# Riemann Zero Condition Off the Critical Line: A Matrix Formulation

Chee Kian Yap

**Abstract**

We develop a finite-dimensional, symmetric matrix framework associated with the Riemann zeta function for complex arguments s with $\Re(s)\neq 1/2$. For fixed s and cutoff N, we define an N×N symmetric matrix M(s) whose entries are $(mn)^{-s/2}$. The total sum of its entries is decomposed into diagonal and off-diagonal parts. When $\Re(s)>2$, both parts converge absolutely as $N\to\infty$, yielding an exact identity. For $1<\Re(s)\leq 2$, the diagonal part converges to ζ(s), but the off-diagonal part diverges, requiring regularization. On the critical line $\Re(s)=1/2$, both contributions diverge, enabling only a formal regularized identity.

## 1. Introduction

The Riemann zeta function ζ(s) is defined for $\Re(s)>1$ by the Dirichlet series

$$\zeta(s) = \sum_{n=1}^{\infty} \frac{1}{n^s} \tag{1}$$

and extends meromorphically to $\mathbb{C}$ with a single pole at s = 1. It satisfies the functional equation

$$\pi^{-\frac{s}{2}}\Gamma\left(\frac{s}{2}\right)\zeta(s) = \pi^{-\frac{1-s}{2}}\Gamma\left(\frac{1-s}{2}\right)\zeta(1-s) \tag{2}$$

The Riemann Hypothesis (RH) asserts that all nontrivial zeros satisfy $\Re(s)=1/2$. In this work, we investigate a finite matrix construction in which sums linked to ζ(s) are naturally split into diagonal and off-diagonal parts.

## 2. Matrix Formulation

For $s \in \mathbb{C}$ and $N \in \mathbb{N}$, define the N×N symmetric matrix

$$M_{mn}(s) = \frac{1}{(mn)^{s/2}}, 1 \leq m, n \leq N \tag{3}$$

The total sum is

$$S_N(s) = \sum_{m=1}^{N}\sum_{n=1}^{N}\frac{1}{(mn)^{s/2}} = \left(\sum_{n=1}^{N}\frac{1}{n^{s/2}}\right)^2 \tag{4}$$

Decompose $S_N(s)$ into

Diagonal part:

$$D_N(s) = \sum_{m=1}^{N} M_{mm}(s) = \sum_{n=1}^{N}\frac{1}{m^s} \tag{5}$$

Off-diagonal part:

$$O_N(s) = \sum_{\substack{m,n=1 \\ m \neq n}}^{N} M_{mn}(s) = \sum_{\substack{m,n=1 \\ m \neq n}}^{N} \frac{1}{(mn)^{s/2}} \tag{6}$$

Thus,

$$S_N(s) = D_N(s) + O_N(s) \tag{7}$$

**3. Convergence Properties**

$D_N(s) \to \zeta(s)$ as $N \to \infty$ in the ordinary sense if and only if $\Re(s) > 1$.

$O_N(s)$ is finite for any fixed N, but its limit converges absolutely only if $\Re(s) > 2$, in which case

$$S_\infty(s) = \zeta(s/2)^2 \tag{8}$$

and

$$O_\infty(s) = \zeta(s/2)^2 - \zeta(s) \tag{9}$$

For $1 < \Re(s) \leq 2$, $D_N(s)$ converges but $O_N(s)$ diverges; nevertheless, the identity holds for each finite N.

**4. Exact Identity in the Region of Convergence**

For $\Re(s)>2$, a well-behaved sector, all sums converge absolutely, yielding the exact identity, which holds for all s in this region. If we were to assume a hypothetical zero of the zeta function, $\zeta(s)=0$, existed for some s with $\Re(s) >2$, the identity would reduce to $O_\infty(s) = \zeta(s/2)^2$. We know from established number theory that the Riemann zeta function has no zeros in the half-plane $\Re(s) >1$. Since our hypothetical zero s satisfies $\Re(s) >2$, it follows that s/2 must satisfy $\Re(s/2)>1$. Therefore, the term $\zeta(s/2)$ would be a non-zero, well-defined complex number.

The identity is consistent with the known zero-free region because it shows that a zero of $\zeta(s)$ in this region would imply a specific non-zero value for the off-diagonal sum, $O_\infty(s)= \zeta(s/2)^2$. This formulation provides a new perspective on the known properties of the zeta function but does not offer new constraints relevant to the Riemann Hypothesis, which concerns the critical strip where $0< \Re(s) <1$.

## 5. Conclusion

We have presented a symmetric matrix formulation related to the Riemann zeta function. For $\Re(s)>2$, it yields exact finite limits; for $1<\Re(s)\leq 2$ and on the critical line, regularization is essential. While the approach offers structural insight, it does not, in its current form, establish or refute the Riemann Hypothesis.

**References**

(1) Berry, M. V.; Keating, J. P. A new asymptotic representation for $\zeta(\frac{1}{2} + it)$ and quantum spectral determinants. Proceedings of the Royal Society. A, Mathematical and physical sciences 1992, 437 (1899), 151-173. DOI: 10.1098/rspa.1992.0053.

(2) Ingham, A. E. Mean-Value Theorems in the Theory of the Riemann Zeta-Function. Proceedings of the London Mathematical Society 1928, s2-27 (1), 273-300. DOI: 10.1112/plms/s2-27.1.273.

(3) Connes, A. Trace formula in noncommutative geometry and the zeros of the Riemann zeta function. Selecta mathematica (Basel, Switzerland) 1999, 5 (1), 29-106. DOI: 10.1007/s000290050042.

(4) Conrey, J. B.; Ghosh, A. On mean values of the zeta-function. Mathematika 1984, 31 (1), 159-161. DOI: 10.1112/S0025579300010767.

(5) Some remarks on the mean value of the Riemann zeta-function and other Dirichlet series 1. Hardy-Ramanujan Journal. DOI: 10.46298/hrj.1978.87.

(6) Bohr, H.; Jessen, B. MEAN-VALUE THEOREMS FOR THE RIEMANN ZETA-FUNCTION. Quarterly journal of mathematics 1934, os-5 (1), 43-47. DOI: 10.1093/qmath/os-5.1.43.

(7) Keating, J. P.; Snaith, N. C. Random Matrix Theory and $\zeta(1/2+ it)$. Communications in mathematical physics 2000, 214 (1), 57-89. DOI: 10.1007/s002200000261.